\documentstyle[prb,preprint,aps,epsf]{revtex}
\begin{document}
\draft
\title{Non-coordinate time/velocity pairs in special relativity}
\author{P. Fraundorf}
\address{Department of Physics \& Astronomy \\
University of Missouri-StL, \\
St. Louis MO 63121}
\date{\today }
\maketitle

\begin{abstract}
Motions with respect to one inertial (or ``map'') frame are often described
in terms of the coordinate time/velocity pair (or ``kinematic'') of the map
frame itself. Since not all observers experience time in the same way, other
time/velocity pairs describe map-frame trajectories as well. Such coexisting
kinematics provide alternate variables to describe acceleration. We outline
a general strategy to examine these. For example, Galileo's acceleration
equations describe unidirectional relativistic motion {\it exactly} if one
uses $V=\frac{dx}{dT}$, where $x$ is map-frame position and $T$ is clock
time in a chase plane moving such that $\gamma ^{\prime }=\gamma \sqrt{\frac 
2{\gamma +1}}$. Velocity in the traveler's kinematic, on the other hand,
has dynamical and transformational properties which were lost by
coordinate-velocity in the transition to Minkowski space-time. Its repeated
appearance with coordinate time, when expressing relationships in simplest
form, suggests complementarity between traveler and coordinate
kinematic views.
\end{abstract}


\eject

\section{Introduction}

Velocity and time, relative to a given inertial frame in special relativity,
is ordinarily described with reference to the behavior of clocks at rest in
the inertial frame of interest. We refer to this as the coordinate
``kinematic'' or time/velocity pair. Of course clocks not at rest in that
reference or ``map'' frame will observe that the same motions, with respect
to the same map, unfold {\it in time} quite differently. This paper is
simply a look at the accelerated motion of a traveling object or traveler,
according to the time elapsed on clocks {\it other than those at rest} in
the inertial map frame with respect to which distances are being measured.
Examples here involve clocks whose speed, relative to the map frame, is a
function solely of the speed of the traveler, although other
``non-coordinate kinematics'' may be of use as well.

The most fundamental example of a non-coordinate kinematic centers about
time elapsed on the clocks of the traveling object. The time used in
describing map motions is thus proper time $\tau $ elapsed along the world
line of the traveler, while velocity in the traveler-kinematic, $u=dx/d\tau $%
, is the spatial component of 4-velocity with respect to the map frame.
Unlike coordinate-velocity, traveler-kinematic velocity $u$ is thus
proportional to momentum, and obeys elegant addition/transformation rules.
In spite of its intuitive simplicity and usefulness for solving
problems in the context of a single inertial frame, discussion of this
quantity (also called ``proper speed'') has been with two 
exceptions\cite{SearsBrehme,Shurcliff} almost 
non-existent\cite{Shurcliff} in basic texts. The equations
for constant proper acceleration look quite different in the traveler
kinematic, although they too have been rarely discussed (except for 
example in the first edition of the classic text by Taylor and 
Wheeler\cite{TaylorWheeler}). Even in that case, they were not 
discussed as part of a time-velocity pair defined in the context of 
one inertial frame.

Another interesting kinematic for describing motion, with respect to a
map-frame, centers about time elapsed on the clocks of a ``Galilean chase
plane''. For unidirectional motion, Galileo's equations for
constant acceleration {\it exactly} describe relativistic motions for our
traveler in such a kinematic. Therefore, with appropriate inter-kinematic 
conversions in hand, Galileo's equations may
offer the simplest path to some relativistic answers! Galilean
time and Galilean-kinematic velocity remain well defined in (3+1)D special
relativity, with kinetic energy held equal to half mass times velocity
squared. For non-unidirectional motion, however, the constant acceleration
equations containing Galilean-observer time look more like elliptic
integrals than like Galileo's original equations.

Since all kinematics share common distance measures in the context of one
inertial reference frame, they allow one to construct x-tv plots showing all
variables in the context of that frame, including universal constant proper
acceleration surfaces on which {\it all flat space-time problems} of this
sort plot. The single frame character of these non-coordinate kinematics
facilitates their graphical and quantitative use in pre-transform relativity
as well, i.e. by students not yet ready for multiple inertial frames\cite
{PreTransform}.

\section{Non-coordinate kinematics in general}

First a word about notation. Upper case $T$ and $V$ are used 
for time and velocity, respectively, in an arbitrary kinematic. Lower 
case $t$ and $v$ have historically been used with both Galilean time,
and with coordinate-time in Minkowski 4-space.  Hence $b$ and $w\equiv 
\frac{dx}{cb}$ are used for the latter time and velocity, to minimize 
confusion. The
traveler-kinematic time/velocity pair is represented by the proper time 
$\tau $ elapsed along the world line of a traveler, and by spatial four-vector
velocity $u\equiv \frac{dx}{d\tau }$. We have also chosen to use physical
rather than dimensionless variables (albeit often arranged in
dimensionless form) to make the equations useful to as wide a readership as
possible.

This paper is basically about three points of view: that of the map, the
traveler, and the observer. All un-primed variables refer to traveler
motion, and all distances are measured in the reference inertial coordinate (%
$x$,$y$) frame. In short, everyone is talking about the traveler, and
using the same map! However, {\it traveler motion on the map} is 
described variously:
(i) from the map point of view with coordinate time $b$ and velocity $w$,
(ii) from the traveler point of view with traveler time $\tau $ and traveler
kinematic velocity $u$, and (iii) from the observer or ``chase plane'' point
of view with observer-time $T$ and observer-kinematic velocity $V\equiv 
\frac{dx}{dT}$. The velocity definitions used here are summarized in Table 
\ref{Table1}.

The most direct way to specify a general observer kinematic might be via the
rate $h$ at which coordinate-time $b$ in the inertial map frame passes per
unit time $T$ in the kinematic of interest. If we associate this observer
time with physical clock time elapsed along the world line of a chase plane
(primed frame) carrying the clocks of that kinematic, then this rate for a
given kinematic can be written

\begin{equation}
h\equiv \frac{db}{dT}=\frac{db}{d\tau ^{\prime }}=\gamma ^{\prime }\geq 1.
\label{h_eqn}
\end{equation}
Since physical $\gamma =\frac{db}{d\tau }$ values are never less than $1$,
it follows that any non-coordinate time $T$, associated with a physical
clock moving with respect to the map frame, {\it will be passing more slowly
than coordinate-time }$b$.

One can also relate this quantity $h$ to velocity $V$ in the kinematic of
interest. Since $V\equiv \frac{dx}{dT}$ and $w\equiv \frac{dx}{db}$, it
follows that the observer-kinematic velocity $V$ of our traveling object,
with respect to the map frame, is related to coordinate-velocity $w$ by $V=wh
$. Since $h\geq 1$ with equality iff $V$ equals $w$, any non-coordinate
velocity $V$, which describes the traveler motion with respect to a map
frame using times from a physical (albeit non-coordinate) clock, {\it will
be higher than the coordinate-velocity }$w$ describing that same motion.

Time/velocity pairs (kinematics) like the ones examined here may be defined
by specifying $h$ as a function of the speed of the traveler (e.g. as a
function of coordinate-velocity $w$). In this way the coordinate-kinematic
itself is specified by setting $h$ to $1$. If we set $h$ equal to $\frac 1{%
\sqrt{1-(\frac wc)^2}}=\gamma $ or $\frac{db}{d\tau }$, where $c$ is the
coordinate-speed of light, then the resulting kinematic is that associated
with the time $\tau $ read by clocks moving with the traveler.

In order to determine $h$ for a Galilean observer, we explore
further the relationship between $h$ and energy factor $\gamma \equiv \frac E%
{mc^2}=1+\frac K{mc^2}=\frac{db}{d\tau }$, where $E$ is total relativistic
energy for our traveler, $K$ is kinetic energy, and $m$ is traveler rest
mass. In particular, given the function $\gamma [V]$, one can obtain $%
h[V]$ from the relation

\begin{equation}
h\equiv \frac{db}{dT}=\frac Vw=\frac V{\sqrt{1-(1/\gamma )^2}}.
\label{h[V]eqn}
\end{equation}
By inverting $\gamma [V]$ to get $V[\gamma ]$, one may then substitute this 
into (%
\ref{h[V]eqn}) to get an expression for $h[\gamma ]$.

If we define the Galilean kinematic as one for which kinetic energy is $K=%
\frac 12mV^2$, then $E=mc^2+\frac 12mV^2$, and hence $\gamma [V]=\frac E{mc^2%
}=1+\frac 12(\frac Vc)^2$. Then $V[\gamma ]=c\sqrt{2(\gamma -1)}$, which
substituted into (\ref{h[V]eqn}) above gives $h=\gamma \sqrt{\frac 2{\gamma
+1}}$. Hence a ``chase plane'', which holds its own $\gamma ^{\prime }$
value at $\gamma \sqrt{\frac 2{\gamma +1}}$ in terms of $\gamma $ for the
traveler, will observe a velocity $V$ which is related to kinetic energy of $%
K=\frac 12mV^2$, for the traveler in the reference inertial (map) frame.
Note that $\gamma ^{\prime }\leq \gamma $, which means that the Galilean
chase plane always moves a bit more slowly than does the traveler. One can
also translate these ``chase plane instructions'' into coordinate-velocity
terms, by replacing $\gamma $ values above with the corresponding $\frac 1{%
\sqrt{1-(\frac wc)^2}}$ expression, and then solving for $w^{\prime }$.

The generic observer-kinematic relationships discussed here, along with
derived values for coordinate, traveler, and Galilean kinematics, are also
listed in Table \ref{Table1}. What follows is a look at the equations
provided by each of these kinematics for describing map trajectories for our
traveler under constant proper acceleration $\alpha $.

\section{Constant proper acceleration}

The frame-invariant proper acceleration is most simply described with some
definitions in 4-vector notation. If one writes the traveler 4-vector 
position
as ${\bf X}=\left\{ {cb,x,y,z}\right\} $, then traveler 4-velocity is ${\bf U%
}=\frac{d{\bf X}}{d\tau }=\left\{ {c}\gamma {,u}_{{x}}{,u}_{{y}}{,u}_{{z}%
}\right\} $, and the 4-acceleration is ${\bf A}=\frac{d{\bf U}}{d\tau }=%
\frac 1m\left\{ \frac 1c\frac{{dE}}{d\tau }{,F}_{{x}}{,F}_{{y}}{,F}_{{z}%
}\right\} $ where $F_i\equiv \frac{dp_i}{d\tau }$ denotes the proper force
(traveler-time derivative of traveler momentum) in the $i$ direction, and $E$
total relativistic energy, with respect to the map frame. Flat space-time
dot products, formed by subtracting the sum of squares of the last three
components from the square of the first component, include $d{\bf X}\bullet d%
{\bf X}=c^2db^2-dx^2-dy^2-dz^2=d\tau ^2$, ${\bf U}\bullet {\bf U}=c^2$, $%
{\bf A}\bullet {\bf U}=0$, and ${\bf A\bullet A}=-\alpha ^2$, where $\alpha $
is the traveler's {\it proper acceleration}.

With no loss of generality, let's define $x$ as the ''direction of proper
acceleration''. In other words, choose $x$ as the direction in which
coordinate-velocity is changing. If the motion is not unidirectional, then
there will also be a component of coordinate-velocity which is not changing.
Choose the $y$-direction as the direction of this
non-changing component of coordinate-velocity, namely $w_y$.
Holding $\alpha $ constant yields a second order differential equation
whose solution, for times and distances measured from the ``turnabout
point'' in the trajectory at which $w_x\Rightarrow 0$, is

\begin{equation}
{\bf X}=\left( \frac{c^2}\alpha \right) \left\{ 
\begin{array}{c}
\gamma _{{y}}{\sinh [}\frac{{\alpha \tau }}c{]} \\ 
\cosh [\frac{{\alpha \tau }}c{]} \\ 
\frac{w_{{y}}}c{\gamma }_{{y}}{\sinh [}\frac{{\alpha \tau }}c{]} \\ 
0
\end{array}
\right\}   \label{Xeqn}
\end{equation}
where the constant $\gamma _y\equiv 1/\sqrt{1-(\frac{w_y}c)^2}$. The
correctness of this solution, as well as a number of intermediate
quantities, can be established by substituting (\ref{Xeqn}) into the
defining equations above. For unidirectional motion, the solution simplifies
to ${\bf X}=\frac{c^2}\alpha \left\{ {\sinh [}\frac{{\alpha \tau }}c{],\cosh
[}\frac{{\alpha \tau }}c{],0,0}\right\} ${.}

A first integral of the motion, which is independent of kinematic, is
helpful for exploring {\it intra-kinematic} solutions of the constant
acceleration equation. Since one thing in common to all kinematics are map 
frame distances and expressions for $\gamma $, we seek a relation between
initial/final $\gamma $ and initial/final position. This is nothing more
than the work-energy theorem for constant proper acceleration, which takes
the form $\gamma _ym\alpha \Delta x=\Delta E=mc^2\Delta \gamma $. For the
special case when times and distances are measured from the turnabout point,
where $\gamma =\gamma _y$ and $x=0$, this says that $\frac \gamma {\gamma _y}%
=1+\frac{\alpha x}{c^2}$.

\section{The examples}

For the coordinate-kinematic case, $V=w$ and $T=b$, so that putting $\gamma
[V]$ from Table \ref{Table1} above into the work-energy theorem above
provides a constant acceleration equation relating $w$ to $x$. It also
yields a first order differential equation in terms of $x$ and $b$, from
which equations relating $b$ to $x$, and $b$ to $w$, can be obtained. The
results for unidirectional motion, along with similarly obtained results for
traveler and Galilean kinematics, are listed in Table \ref{Table2}. Note in
particular that the simplest result is found using the Galilean kinematic,
wherein the full and familiar set of Galilean constant acceleration
equations become available for the solving of relativistic acceleration
problems. The usefulness of this fact depends on the availability of simple
conversions {\it between kinematics}, to be discussed in the section below.

If motion is not unidirectional, the strategy above is complicated by the
fact that $V$ is not $\frac{dx}{dT}$, but instead is connected to $\frac{dx}{%
dT}$ through the relation $V^2=\sqrt{(\frac{dx}{dT})^2+(hw_y)^2}$, where $h$
may also be a function of $V$. The strategy above still yields a first order
equation in terms of $x$ and $T$, with only minor resulting complication in
the coordinate and traveler kinematic cases. The results of this for (3+1)D
motion, in the three example kinematics, are summarized in Table \ref{Table3}%
.

As you can see from the Table on (3+1)D acceleration, all 14 variables ($x$, 
$y$, $\gamma $, $b$, $w$, $w_x$, $\tau $, $u$, $u_x$, $u_y$, $t$, $v$, $v_x$%
, $v_y$) and 2 constants ($\alpha $, $w_y$) are simply related, except for
Galilean time. We here take advantage of the fact that Galilean time $t$ is
directly expressable in terms of traveler time $\tau $, through the
dimensionless expression 
\begin{equation}
\frac{\alpha t}c=\int \sqrt{\frac{\gamma +1}2}\frac{\alpha d\tau }c=-i\sqrt{%
2(1+\gamma _y)}E[\frac i2\frac{\alpha \tau }c|\frac{2\gamma _y}{(1+\gamma _y)%
}]\equiv \Xi [\frac{\alpha \tau }c]  \label{t_eqn}
\end{equation}
where $E[\phi |m]$ is the standard elliptic integral of the second kind. The
function $\Xi [\frac{\alpha \tau }c]=\frac{\alpha t}c$ defined above, and
it's inverse, $\Xi ^{-1}[\frac{\alpha t}c]=\frac{\alpha \tau }c$, allows
table entries involving Galilean time to be shortened considerably.

\section{Inter-kinematic relations}

The connection between kinematics defined in the context of a single
inertial frame, like the three example kinematics above, can be described
graphically as well as with equations. The velocity inter-conversions can be
obtained directly from Table \ref{Table1} above. For example, velocity 1 can
be used to calculate velocity 2 using $V_2[\gamma [V_1]]$. Time increments
during a period of constant proper acceleration $\alpha $, on the other
hand, are more complicated. For the example kinematics described here, they
seem to be most simply expressed in terms of $\alpha $ and a pair of initial
and final velocities. For the unidirectional motion case, the simplest {\it %
complete set} of such inter conversions between the three example kinematics
is provided in Table \ref{Table4}. Similar conversions for traveler and
coordinate-kinematics, in the case of (3+1)D motion, are provided in Table 
\ref{Table5}.

As a matter of practice when dealing with multiple time and velocity types,
all defined within context of a single inertial frame, it is helpful to
append the kinematic name to the time units, to avoid confusion. Thus, time
in the traveler-kinematic may be measured in [traveler years], while
velocities in the traveler-kinematic are measured in [light-years per
traveler year]. Thus, from Table 4, a traveler-kinematic velocity of $u=1$%
[light-year per traveler year] corresponds to a coordinate-kinematic
velocity of $w=\frac 1{\sqrt{2}}$[$c$], where of course [$c$] is shorthand
for a [light-year per coordinate year]. This unitary value of
traveler-kinematic velocity may serve as a landmark in the transition
between non-relativistic and relativistic behaviors. For these reasons, we
expect that a shorthand units notation for traveler-kinematic velocity, in
particular for a [light-year per traveler year], may prove helpful in days
ahead.

Before leaving the acceleration equations, it is worthwhile summarizing the
findings, in mixed kinematic form as integrated expressions for constant
proper acceleration. For unidirectional motion, or (1+1)D special
relativity, we have shown that

\begin{equation}
\alpha =c^2\frac{\Delta \gamma }{\Delta x}=\frac{\Delta u}{\Delta b}=c\frac{%
\Delta \eta }{\Delta \tau }=\frac{\Delta v}{\Delta t},  \label{a1D}
\end{equation}
where rapidity $\eta \equiv \sinh ^{-1}[\frac uc],$ and lower case $v$ and $t
$ refer to Galilean kinematic variables. In the (3+1)D case, this equation
becomes

\begin{equation}
\alpha =\left( \frac{c^2}{\gamma _y}\right) \frac{\Delta \gamma }{\Delta x}%
=\gamma _y\frac{\Delta u_x}{\Delta b}=c\frac{\Delta \eta _x}{\Delta \tau }=c%
\frac{\Delta \Xi }{\Delta t},  \label{a3D}
\end{equation}
where $\eta _x\equiv \sinh ^{-1}[\frac{u_x}c]$, and $\Xi $ is defined in
equation (\ref{t_eqn}) above.

\begin{figure}[tbp]
\centerline{\epsfysize=7.0in\epsfbox{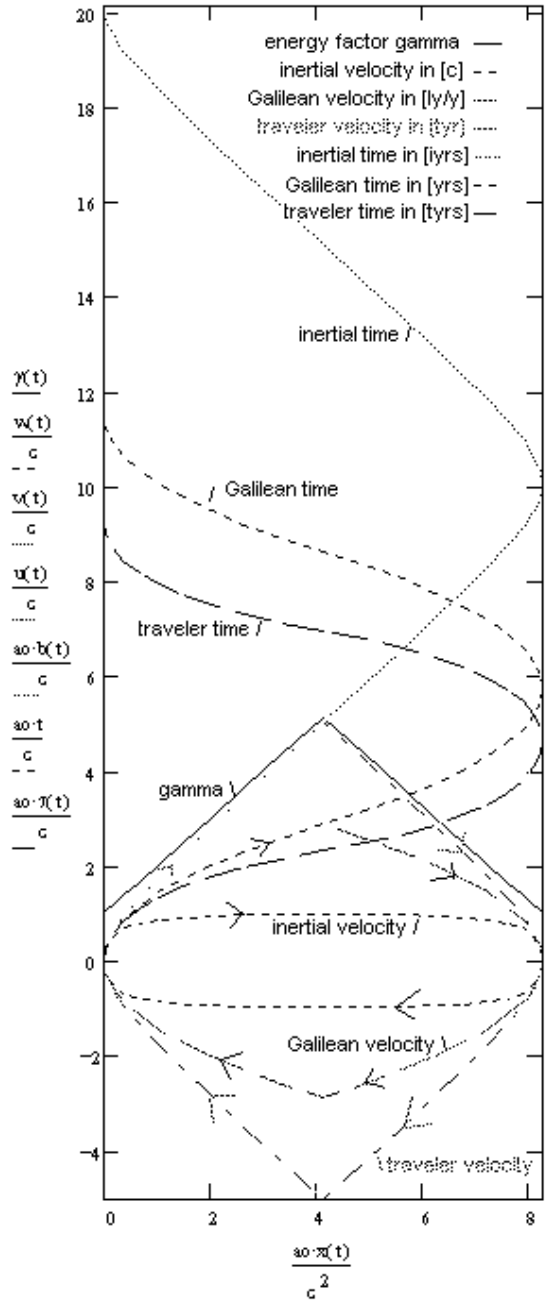}}
\caption{Plot of three kinematic time/velocity pairs, and gamma,
during a 4-part uniformly accelerated round-trip in (1+1)D spacetime. Note
that in the context of simultaneity experienced in the coordinate (map)
frame, concurrent values for all 7 kinematic parameters, at all points in
the traveler's trajectory, are visible at a glance.}
\label{Fig1}
\end{figure}

All of these velocities and times, for a given traveler with respect to a
given map frame, may be plotted as trajectories over the x,y coordinate
field. If we divide velocities by $c$ and times by $\frac c\alpha $, they
can all be plotted on the same graph! For unidirectional motion, this allows
us (for example) to plot all of these parameters as a function of position
even if the acceleration changes, as in the 4-segment twin adventure
illustrated in Figure \ref{Fig1}.

\begin{figure}[tbp]
\centerline{\epsfysize=7.0in\epsfbox{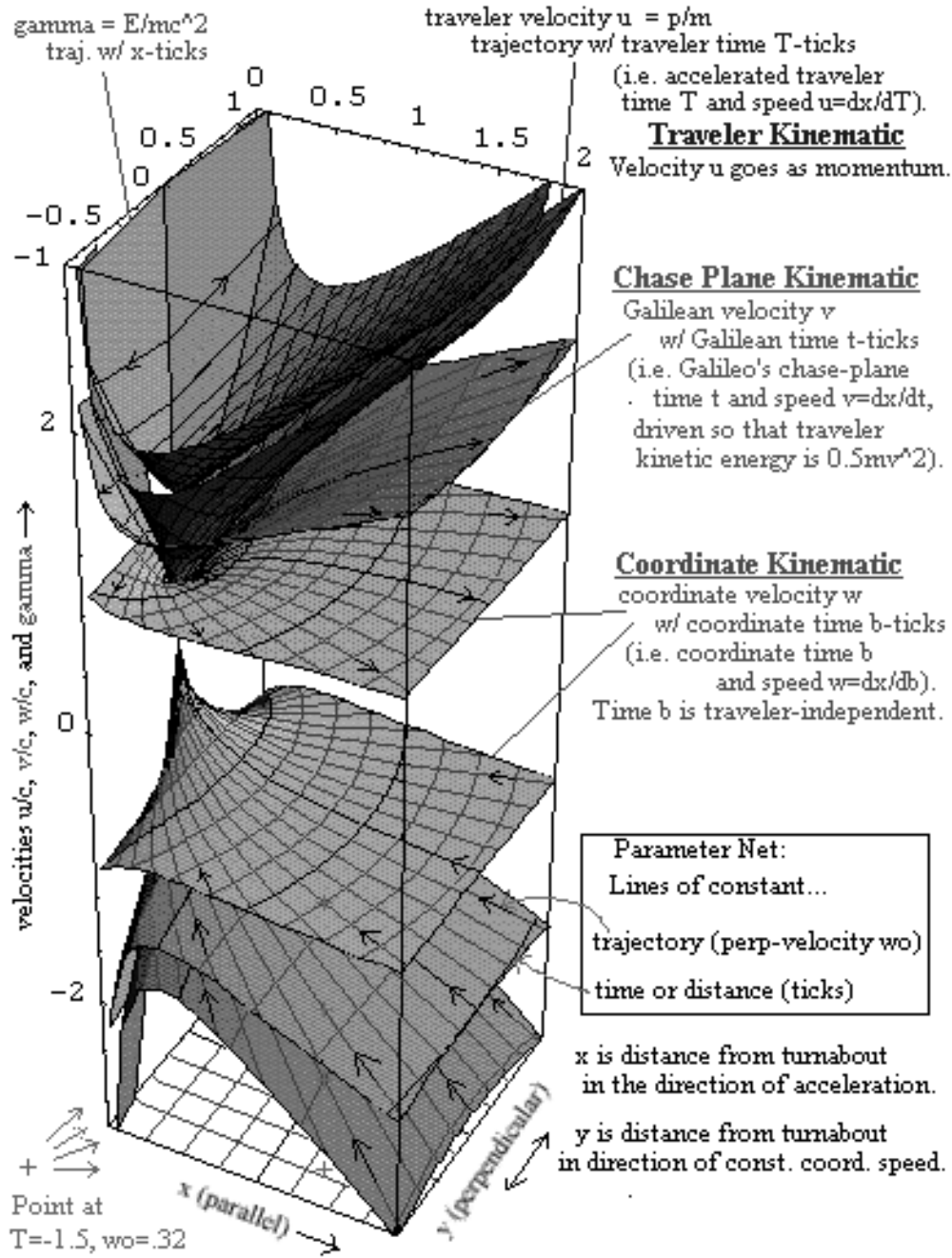}}
\caption{Plot of gamma, as well as of coordinate, traveler, and
Galilean speed, versus position $x$,$y$ measured in an inertial reference
frame, for all uniformly accelerated trajectories in (3+1)D spacetime. That
half of the trajectories which approach turnabout are plotted on the
''inverted flower'' below, even though the speeds being plotted are never
less than zero. Each surface is parameterized with equally spaced values of
time {\it in the kinematic of the speed being plotted}, as well as in
equally spaced values for constant transverse coordinate velocity $w_y$.}
\label{Fig2}
\end{figure}

If we further divide distances in the x,y coordinate field by $\frac{c^2}%
\alpha $, and restrict trajectories to those involving a single constant
acceleration trajectory, or to a family of such trajectories in (3+1)D, a
single universal acceleration plot in the velocity range of interest can be
assembled, on which all constant acceleration problems plot! A plot of this
sort, of the three example kinematic velocities and gamma, over the linear
velocity range from 0 to 3 [light-years per traveler year], is provided in
Figure \ref{Fig2}. Here the various velocity surfaces have been
parameterized using equal time increments in the kinematic of the velocity
being plotted, as well as with equally-spaced values of the constant
transverse coordinate-velocity $w_y$. It is thus easy to see that coordinate
time during constant acceleration passes more quickly, as a function of
position, than do either of the two kinematic times, as predicted more
generally by equation (\ref{h_eqn}).

All of the foregoing has been done in the context of a single inertial
frame. The transformation properties of these various times and velocities
between frames also deserve mention. As is well known, and apparent from the
section on 4-vector acceleration above, {\it coordinate-time} is the time
component of the coordinate 4-vector, while {\it traveler-kinematic velocity}
provides spatial components for the velocity 4-vector. {\it %
Coordinate-velocity} enjoys no such role, and its comparatively messy
behaviors when transforming between frames need no further introduction.

A similar inelegance is found in the well-known rule of velocity addition
for coordinate-velocities. A comprehensive look at addition rules for the
velocities used here, and for $\gamma $ as well, shows that the simplest of
all (to this author at least) is the rule for the relative
traveler-kinematic velocity associated with particles A and B on a collision
course, namely $u_{rel}=\gamma _A\gamma _B(w_A+w_B)$. Recall that within a
single frame, $u=\gamma w$. Hence on addition, the $\gamma $ factors of
traveler-kinematic velocity multiply, while the coordinate-velocity factors
add.

\section{Discussion}

The coordinate-kinematic has been used almost exclusively in the special
relativity literature to date. This is in part because the {\it %
traveler-independent nature of time} in a world of Galilean transforms finds
its strongest (if still weak) natural analog in the coordinate-kinematic of
Minkowski space-time. However, other features of velocity and time abandon
the coordinate-kinematic, when the transition from Galilean to Lorentz
transforms is made.

For example, we lose the second time derivative of coordinate position as a
useful acceleration, and thus the coordinate kinematic loses {\it Galileo's
equations for constant acceleration}. These equations remain intact for
unidirectional motion if we put our clocks into a Galilean chase plane. The
role of this kinematic, in allowing proper acceleration to be written as a
second time derivative of position (i.e. for which $V=\alpha T$), is
illuminated by noting that unidirectional motion from rest is described by $%
\frac{\alpha x}{c^2}=\cosh [\frac{\alpha \tau }c]-1$. Hence $\frac uc=\frac{%
\alpha b}c=\sinh [\frac{\alpha \tau }c]$, but $u$ and $b$ are in different
kinematics. If we want $\frac Vc=\frac{\alpha T}c$ for time/velocity in one
kinematic, a $V[\tau ]$ obeying $\alpha dx=\alpha VdT=VdV=\alpha c\sinh [%
\frac{\alpha \tau }c]d\tau $ is required. One can rewrite the right side of
this equation as $2c\sinh [\frac{\alpha \tau }{2c}]$ times $\alpha \cosh [%
\frac{\alpha \tau }{2c}]d\tau =d[2c\sinh [\frac{\alpha \tau }{2c}]]$, or as $%
VdV$ if and only if $\frac Vc=2\sinh [\frac{\alpha \tau }{2c}]=\sqrt{2(\frac{%
\alpha x}{c^2})}$. Thus Galilean velocity is the only way to make proper
acceleration a second time derivative in (1+1)D. From similarities between $%
\frac vc=2\sinh [\frac{\alpha \tau }{2c}]$ and the expression above for $%
\frac uc=\frac{\alpha b}c$, it is easy to see why the Galilean kinematic
provides an excellent low-velocity (small-argument) approximation to
traveler velocity and coordinate time.

On a deeper note, {\it the inertial nature of velocity} in Galilean
space-time (i.e. its proportionality to momentum), as well as its relatively 
{\it elegant transformation properties}, do not belong to coordinate
velocity in Minkowski space-time. Velocity in the traveler kinematic
inherits them instead. The traveler-kinematic also has relatively {\it %
simple equations for constant acceleration}. However, constant acceleration
is most simply expressed in mixed-kinematic terms, using the coordinate-time
derivative of a traveler-kinematic velocity component.

Thus we have a choice, to cast expressions which involve velocity in terms
of {\it either} the coordinate-time derivative of position ($w$), {\it or}
in terms of the dynamically and transformationally more robust traveler-time
derivative of position ($u$). Some of the connections discussed here might
be more clearly reflected than they are at present in the language that we
use, had historical strategies taken the alternate path.

\acknowledgments

This work has benefited directly from support by UM-St.Louis, and
indirectly from support by the U.S. Department of Energy, the 
Missouri Research Board, as well as Monsanto and MEMC Electronic 
Materials Companies. Thanks also to family, friends, and colleagues 
{\it at all stages in their careers}, for their thoughts 
and patience concerning this ``little project on the side''.

\begin{table}[tbp] \centering
\caption{Defining multiple kinematics in context of one inertial
frame\label{Table1}}%
\begin{tabular}{rcccc}
{\bf kinematic:} & generic $T$,$V$ & coordinate $b$,$w$ & traveler $\tau $,$%
u $ & Galilean $t$,$v$ \\ \hline
velocity of traveler & $V\equiv \frac{dx}{dT}$ & $w\equiv \frac{dx}{db}$ & $%
u\equiv \frac{dx}{d\tau }$ & $v\equiv \frac{dx}{dt}$ \\ 
$\frac E{mc^2}=\gamma [V]=$ & \{given\} & $\frac 1{\sqrt{1-(w/c)^2}}$ & $%
\sqrt{1+(u/c)^2}$ & $1+\frac 12(v/c)^2$ \\ 
$\gamma ^{\prime }=\frac{db}{dT}=h[V]=$ & $\frac{V/c}{\sqrt{1-1/\gamma [V]^2}%
}$ & $1$ & $\sqrt{1+(u/c)^2}$ & $\frac{v/c}{\sqrt{1-(1+\frac 12(v/c)^2)^{-2}}%
}$ \\ 
$V[\gamma ]=$ & \{invert $\gamma [V]$\} & $c\sqrt{1-\frac 1{\gamma ^2}}$ & $c%
\sqrt{\gamma ^2-1}$ & $c\sqrt{2(\gamma -1)}$ \\ 
$\gamma ^{\prime }=\frac{db}{dT}=h[\gamma ]=$ & $h/V[\gamma ]]$ & $1$ & $%
\gamma $ & $\gamma \sqrt{\frac 2{\gamma +1}}$ \\ 
$\frac{\gamma ^{\prime }}\gamma =\frac{dT}{d\tau }[\gamma ]=$ & $\frac \gamma
{h[\gamma ]}$ & $\gamma $ & $1$ & $\sqrt{\frac{\gamma +1}2}$%
\end{tabular}
\end{table}

\begin{table}[tbp] \centering
\caption{Uni-directional constant acceleration from rest in 3
kinematics\label{Table2}}%
\begin{tabular}{rccc}
{\bf kinematic:} & coordinate & traveler & Galilean \\ \hline
speed $V[x]$ & $\frac wc=\sqrt{1-\frac 1{\left( 1+\frac{\alpha x}{c^2}%
\right) ^2}}$ & $\frac uc=\sqrt{\left( 1+\frac{\alpha x}{c^2}\right) ^2-1}$
& $\frac vc=\sqrt{2\left( \frac{\alpha x}{c^2}\right) }$ \\ 
distance $x[T]$ & $\frac{\alpha x}{c^2}=\sqrt{1+\left( \frac{\alpha b}c%
\right) ^2}-1$ & $\frac{\alpha x}{c^2}=\cosh [\frac{\alpha \tau }c]-1$ & $%
\frac{\alpha x}{c^2}=\frac 12\left( \frac{\alpha t}c\right) ^2$ \\ 
time $T[V]$ & $\frac{\alpha b}c=\sqrt{\frac 1{\left( \frac wc\right) ^2}-1}$
& $\frac{\alpha \tau }c=\sinh ^{-1}[\frac uc]$ & $\frac{\alpha t}c=\frac vc$%
\end{tabular}
\end{table}

\begin{table}[tbp] \centering
\caption{(3+1)D constant acceleration from turnabout in 3
kinematics\label{Table3}}%
\begin{tabular}{lccc}
{\bf kinematic:} & coordinate & traveler & Galilean \\ \hline
$\frac E{mc^2}$ or $\gamma [x]$ & $\gamma =\gamma _y\left( 1+\frac{\alpha x}{%
c^2}\right) $ & $\gamma =\gamma _y\left( 1+\frac{\alpha x}{c^2}\right) $ & $%
\gamma =\gamma _y\left( 1+\frac{\alpha x}{c^2}\right) $ \\ 
speed $V[\gamma ]$ & $\frac wc=\sqrt{1-\frac 1{\gamma ^2}}$ & $\frac uc=%
\sqrt{\gamma ^2-1}$ & $\frac vc=\sqrt{2(\gamma -1)}$ \\ 
x-velocity $V_x$ & $\frac{w_x}c=\sqrt{\frac{w^2}{c^2}-\frac{w_y^2}{c^2}}$ & $%
\frac{u_x}c=\sqrt{\frac{u^2}{c^2}-\gamma ^2\frac{w_y^2}{c^2}}$ & $\frac{v_x}c%
=\sqrt{\frac{v^2}{c^2}-\frac{2\gamma ^2}{(\gamma +1)}\frac{w_y^2}{c^2}}$ \\ 
y-velocity $V_y$ & \{$\frac{w_y}c$ given, constant\} & $\frac{u_y}c=\gamma 
\frac{w_y}c$ & $\frac{v_y}c=\sqrt{\frac 2{\gamma +1}}\gamma \frac{w_y}c$ \\ 
distance $x[T]$ & $\frac{\alpha x}{c^2}=\sqrt{1+\frac 1{\gamma _y}\left( 
\frac{ab}c\right) ^2}-1$ & $\frac{\alpha x}{c^2}=\cosh [\frac{\alpha \tau }c%
]-1$ & $\frac{\alpha x}{c^2}=\cosh [\Xi ^{-1}[\frac{\alpha t}c]]-1$ \\ 
distance $y[T]$ & $\frac{\alpha y}{c^2}=\frac{w_y}c\frac{\alpha b}c$ & $%
\frac{\alpha y}{c^2}=\frac{w_y}c\gamma _y\sinh [\frac{\alpha \tau }c]$ & $%
\frac{\alpha y}{c^2}=\gamma _y\frac{w_y}c\sinh [\Xi ^{-1}[\frac{\alpha t}c]]$
\\ 
time $T[V_x]$ & $\frac{\alpha b}c=\gamma _y\sqrt{\frac{c^2}{w_x^2}-1}$ & $%
\frac{\alpha \tau }c=\sinh ^{-1}\left[ \frac{u_x}c\right] $ & $\frac{\alpha t%
}c=\Xi [\cosh ^{-1}[\frac{1+\frac 12(\frac vc)^2}{\gamma _y}]]$%
\end{tabular}
\end{table}

\begin{table}[tbp] \centering
\caption{Interkinematic conversions for speed, and uni-directional
acceleration\label{Table4}}%
\begin{tabular}{lccc}
{\bf From}$\rightarrow ${\bf To} & energy factor $\gamma $ & initial/final
speed & (1+1)D elapsed time \\ \hline
Galilean$\rightarrow $Traveler & $\gamma =1+\frac 12\left( \frac vc\right) ^2
$ & $u=v\sqrt{1+\frac 14\left( \frac vc\right) ^2}$ & $\Delta \tau =\frac c%
\alpha \Delta \sinh ^{-1}[\frac uc]$ \\ 
Traveler$\rightarrow $Coord. & $\gamma =\sqrt{1+\left( \frac uc\right) ^2}$
& $w=\frac u\gamma $ & $\Delta b=\frac{\Delta u}\alpha $ \\ 
Coord.$\rightarrow $Galilean & $\gamma =\frac 1{\sqrt{1-\left( \frac wc%
\right) ^2}}$ & $v=\sqrt{\frac 2{\gamma +1}}\gamma w$ & $\Delta t=\frac{%
\Delta v}\alpha $%
\end{tabular}
\end{table}

\begin{table}[tbp] \centering
\caption{Interkinematic conversions for the (3+1)D
case\label{Table5}}%
\begin{tabular}{lccc}
{\bf From}$\rightarrow ${\bf To} & initial/final $V_y$ & initial/final $V_x$
& (3+1)D elapsed time \\ \hline
Galilean$\rightarrow $Traveler & $u_y=\gamma w_y$ & $u_x=\sqrt{u^2-\gamma
^2w_y^2}$ & $\Delta \tau =\frac c\alpha \Delta \sinh ^{-1}[\frac{u_x}c]$ \\ 
Traveler$\rightarrow $Coord. & \{given, as $w_y$\} & $w_x=\sqrt{w^2-w_y^2}$
& $\Delta b=\gamma _y\frac{\Delta u_x}\alpha $ \\ 
Coord.$\rightarrow $Galilean & $v_y=\sqrt{\frac 2{\gamma +1}}\gamma w_y$ & $%
v_x=\sqrt{v^2-\frac{2\gamma ^2w_y^2}{\gamma +1}}$ & $\Delta t=\frac c\alpha
\Delta \Xi [\cosh ^{-1}[\frac \gamma {\gamma _y}]]$%
\end{tabular}
\end{table}

\end{document}